\documentclass[twocolumn,journal]{IEEEtran}
\usepackage[noadjust]{cite}

\ifCLASSINFOpdf
   \usepackage[pdftex]{graphicx}
\else
   \usepackage[dvips]{graphicx}
\fi
\usepackage[colorlinks,
            linkcolor=black,
            urlcolor=red,
            anchorcolor=black,
            citecolor=black
            ]{hyperref}
\usepackage{tabularx}
\usepackage{supertabular}
\usepackage{amsfonts}
\usepackage{bm}
\usepackage{longtable}
\usepackage{xcolor}
\usepackage{lineno}
\usepackage[cmex10]{amsmath}
\usepackage{fixltx2e}
\usepackage{url}

\begin{document}
\title{Construction and Performance of Quantum Burst Error Correction Codes for Correlated Errors}
%

\author{Jihao Fan, Min-Hsiu Hsieh, Hanwu Chen, Yonghui Li, and He Chen
 \thanks{ }
 \thanks{ }
 \thanks{ }
}

\author{
\IEEEauthorblockN{Jihao Fan\IEEEauthorrefmark{1}\IEEEauthorrefmark{3}, Min-Hsiu Hsieh\IEEEauthorrefmark{2}, Hanwu Chen\IEEEauthorrefmark{3}, He Chen\IEEEauthorrefmark{4}, and Yonghui Li\IEEEauthorrefmark{4}}

\IEEEauthorblockA{\IEEEauthorrefmark{1}Nanjing Institute of Technology, Nanjing, Jiangsu, China}

\IEEEauthorblockA{\IEEEauthorrefmark{2}University of Technology Sydney, Australia }

\IEEEauthorblockA{\IEEEauthorrefmark{3}Southeast University, Nanjing, Jiangsu, China}

\IEEEauthorblockA{\IEEEauthorrefmark{4}University of Sydney, Australia}

 \IEEEauthorblockA{jihao.fan@outlook.com, min-hsiu.hsieh@uts.edu.au, hw\_chen@seu.edu.cn, \{he.chen, yonghui.li\}@sydney.edu.au}}

\maketitle

\begin{abstract}
In practical communication and computation systems, errors occur predominantly in adjacent positions rather than in a random manner.  In this paper, we develop a stabilizer formalism for quantum burst error correction codes (QBECC) to combat such error patterns in the quantum regime. Our contributions are as follows. Firstly, we derive an upper bound for the correctable burst errors of QBECCs, the quantum Reiger bound (QRB). This bound generalizes the quantum Singleton bound for standard quantum error correction codes (QECCs). Secondly, we propose two constructions of QBECCs: one by heuristic computer search and the other by concatenating two quantum tensor product codes (QTPCs).  We obtain several new QBECCs with better parameters than existing codes with the same coding length. Moreover, some of the constructed codes can saturate the quantum Reiger bounds. Finally, we perform numerical experiments for our constructed codes over Markovian correlated depolarizing quantum memory channels, and show that QBECCs indeed outperform standard QECCs in this scenario.

\end{abstract}


%
\IEEEpeerreviewmaketitle

\section{Introduction}
\newcounter{conter1}
\setcounter{conter1}{0}
\newtheorem{theorem}[conter1]{Theorem}

\newtheorem{definitions}{Definition}
\newtheorem{theorems}{Theorem}
\newtheorem{lemmas}{Lemma}
\newtheorem{corollarys}{Corollary}
\newtheorem{examples}{Example}
\newtheorem{propositions}{Proposition}
%
%
%
%

In classical information theory, there are mainly two types of error models: the independent noise model proposed by Shannon and the adversarial noise model considered by Hamming. Errors in these two models are usually called random errors and burst errors. Correspondingly, there are random error-correcting codes (RECC) and burst error-correcting codes (BECC) to deal with these two different types of errors \cite{lin2004error}. In reality, channels tend to introduce errors which are localized in a short interval, i.e., the burst errors. These errors could be commonly found in communication systems and storage mediums, as a result of a stroke of lightning in wireless channels or scratch on a storage disc.

In the quantum regime, quantum errors can be independent or correlated in space and time. Hence there are counterparts of  quantum  random error-correcting codes {(see, e.g., \cite{steane1996error,  calderbank1996good, calderbank1997quantum, Brun436, 6671483}) } and quantum burst error-correcting codes {\cite{jihao2017on, vatan1999spatially, kawabata2000quantum, tokiwa2005some}}. Analogous to the classical case, quantum channels commonly have memory \cite{PhysRevA.72.062323} or introduce errors which are localized \cite{caruso2014quantum}, i.e., quantum burst errors.
Vatan \textit{et al.}~\cite{vatan1999spatially}  first considered spatially correlated qubit errors and constructed families of QBECCs  using CSS construction \cite{calderbank1996good, calderbank1997quantum}. However, CSS construction yields QBECCs with inferior code rate. In \cite{kawabata2000quantum},  a quantum interleaver for QBECCs was proposed so that long QBECCs could be produced  from short ones. However, this method highly relies on short efficient QBECCs, which are lacking at this moment.
In \cite{tokiwa2005some}, QBECCs of length up to $51$ were found using computer search.

The construction and investigation of QBECCs have received far less attention, compared to the development of standard QECCs or entanglement-assisted QECCs \cite{7501532, Brun436, 6461941, 6671483, 5714249, PhysRevA.79.032340, PhysRevA.76.062313}. Many important questions remain open. Currently, there is no general upper bound for correctable quantum burst errors, analygous to the classical Reiger bound: $n-k\geq 2\ell$, of an $[n,k]$ classical BECC, where $n$ is the code length, $k$ is the message size, and $\ell$ is the correctable length of burst errors.
In addition,  there is an interesting class of quantum codes, called \emph{degenerate} codes, that have no classical correspondences. They can potentially  store more quantum information or correct more quantum errors than nondegenerate codes. However, degenerate QBECCs have never been explored.

In this paper we  generalize the theory of standard QECCs  to  QBECCs.  We develop the stabilizer formalism for QBECCs and prove the corresponding quantum Reiger bound: $n-k\geq 4\ell$,  for an $[[n,k]]$ QBECC that  corrects a quantum burst error of length $\ell$ or less. The quantum Reiger bound further generalizes the quantum Singleton bound in QECCs.  We obtain many new  QBECCs in the stabilizer formalism    via computer heuristic search, and these codes  are better than existing QBECCs with the same code lengths. We show that the burst error-correcting abilities of most of these codes can achieve the quantum Reiger bound and thus are perfect codes.
In particular, several of our constructed QBECCs, that attain the quantum Reiger bound, are degenerate codes. Additionally, we  propose a new concatenation construction of long QBECCs from two short component codes based on the quantum tensor product code structure \cite{jihao2017on} and the interleaving technique.  Since only one of the component codes of QTPCs needs to satisfy the dual containing constraint, this construction method can largely facilitate the systematical construction of QBECCs. Finally, we perform numerical experiments on two of our constructed QBECCs by
measuring the entanglement fidelity over Markovian correlated depolarizing quantum memory channels \cite{caruso2014quantum}. It is known that the correlation errors in memory channels can lower the
performance of the entanglement fidelity of standard QECCs (see \cite{Cafaro2010quantum,caruso2014quantum}). But if we consider the extra burst error correction abilities of them, they can indeed outperform the best QECCs of the same lengths for random errors.


\section{Theory of Quantum Burst Error Correction Codes}
In this section, we introduce background of QECCs and develop the stabilizer formalism for QBECCs.

\subsection{Quantum Burst Error Correction Codes}

In a two-dimensional complex Hilbert space $\mathbb{C}^2$, a qubit  $|v\rangle$ can be written as
$
|v\rangle=\alpha|0\rangle+\beta|1\rangle,
$
where $\alpha$ and $\beta$ are complex numbers satisfying $|\alpha|^2+|\beta|^2=1$. Two states $|v\rangle$ and $e^{i\theta}|v\rangle$, that are different up to a global phase $e^{i\theta}$, are considered to be the same  in this paper. The Pauli matrices
\begin{equation*}
\label{Pauli_Matrices}
I_2=  \left[
\begin{matrix}
1&0\\
0&1\\
\end{matrix}
\right],
X=  \left[
\begin{matrix}
0&1\\
1&0\\
\end{matrix}
\right],
Z=  \left[
\begin{matrix}
\setlength{\arraycolsep}{0.1pt}
1&0\\
0&-1\\
\end{matrix}
\right],
Y=  \left[
\begin{matrix}
\setlength{\arraycolsep}{0.1pt}
0&-i\\
i&0\\
\end{matrix}
\right]
\end{equation*}
form a basis of the linear operators on $\mathbb{C}^2$, where $i=\sqrt{-1}$ and $Y=iXZ$. An $n$-qubit $|\psi\rangle$ is then a quantum state in the $n$-th tensor product   of $\mathbb{C}^2$, i.e., $|\psi\rangle \in \mathbb{C}^{2^n}\equiv \mathbb{C}^2\otimes\mathbb{C}^2\otimes\cdots\otimes\mathbb{C}^2$.

Since it is possible to \emph{discretize} quantum errors \cite{nielsen2000quantum, 6778074}, we only need to consider a discrete set of quantum errors of $n$ qubits,
described by the following error group
\begin{equation}
\mathcal{G}_n=\{i^\lambda w_1\otimes\cdots\otimes w_n|0\leq\lambda\leq 3, w_i\in{I_2,X,Y,Z}\}.
\end{equation}
Furthermore, it is sufficient to consider the quotient group $\mathcal{\overline{G}}_n=  \mathcal{G}_n/\{\pm1,\pm i\}$ of  $\mathcal{G}_n$ since the global phase  $i^\lambda$ in $\mathcal{G}_n$ is not important. Let $\overline{e}= w_1\otimes w_2\otimes\cdots\otimes w_n\in\mathcal{\overline{G}}_n$ and $e=i^\lambda\overline{e}\in\mathcal{G}_n$.  We define the \emph{burst} length of $\bar{e}$ to be $\ell$, denoted  by $\textrm{bl}_Q(e)=\textrm{bl}_Q(\overline{e})=\ell$, if the nonidentity matrices in  $\overline{e}$ are confined to at most $\ell$ consecutive $w_i$'s.

The idea of a QECC is to encode quantum information into a  subspace of some larger Hilbert space. An $((n,K))$ QECC $Q$ is defined to contain the subspace of dimension $K$ in $\mathbb{C}^{2^n}$. If $K=2^k$, then $Q$ is also written as $Q=[[n,k]]$. A QECC $Q$ can correct  arbitrary errors from an error class $\varepsilon$ \cite{bennett1996mixed,knill1997theory} if
\begin{equation}\label{knill77}
\langle c_i|EE'|c_j\rangle=a_{(E,E')}\delta_{ij}
\end{equation}
for all $\langle c_i|c_j\rangle = \delta_{ij}$ and for all $E\neq E' \in\varepsilon$, where $|c_i\rangle$ and $|c_j\rangle\in Q$, and $a_{(E,E')}$ is a constant     which depends only on $E$ and $E'$. If $\langle c_i|EE'|c_j\rangle=0$ for all $|c_i\rangle,|c_j\rangle\in Q$ and for all $E\neq E' \in\varepsilon$, then $Q$ is called a \emph{nondegenerate} quantum  code.

%

The above error-correcting condition (\ref{knill77}) can be generalized to the burst error case.
\begin{propositions}
\label{proposition_burst}
The code $Q$ can correct any quantum    burst errors of length $l$ or less if and only if
\begin{equation}
\label{burst-error-correction criterion}
\langle c_i|EE'|c_j\rangle=a_{(E,E')}\delta_{ij}
\end{equation}
for all $\langle c_i|c_j\rangle = \delta_{ij}$ and for all $\textrm{bl}(E),\textrm{bl}(E')\leq \ell$, where $|c_i\rangle$ and $|c_j\rangle\in Q$, $E$ and $E'\in\mathcal{G}_n$, and $a_{(E,E')}$ is a constant     which depends only on $E$ and $E'$.

If $\langle c_i|EE'|c_j\rangle=0$ for all $|c_i\rangle,|c_j\rangle\in Q$ and for all $\textrm{bl}(E),\textrm{bl}(E')\leq \ell$, where $E\neq E' \in\mathcal{G}_n$, then $Q$ is a  nondegenerate QBECC.
\end{propositions}

According to the group theoretic framework for QECCs in  \cite{calderbank1997quantum,calderbank1998quantum}, we can also get the stabilizer formalism for QBECCs.
Let $(a|b)$ and $(a'|b')$ be two vectors in $\mathbb{F}_2^{2n}$, the \emph{symplectic inner product} of them is given by
\begin{equation}
 ((a|b),(a'|b'))_s=a\cdot b'+a'\cdot b.
\end{equation}
For a subspace $C$ of   $\mathbb{F}_2^{2n}$, the \emph{symplectic  dual} space  $C^{\bot_s}$ of $C$ is given by
\begin{equation}
 C^{\bot_s}=\{u\in \mathbb{F}_2^{2n}|\forall c\in C,(u,c)_s=0\}.
\end{equation}
We  define  the \emph{symplectic burst}   length of  a nonzero vector $(a|b)=(a_1\cdots a_n|b_1\cdots b_n)\in \mathbb{F}_2^{2n}$ to be the
largest integer  $1\leq \ell\leq n$ such that $(a_i|b_i)\neq (0,0)$ and $(a_{i+\ell-1}|b_{i+\ell-1})\neq (0,0)$ for some $1\leq i\leq n$. We denote by $\textrm{bl}_s((a|b))=\ell$.

According to \cite{calderbank1998quantum}, each element $E\in\mathcal{G}_n$ can be written
uniquely as $ E=i^\lambda X(a)Z(b)$ where $1\leq\lambda\leq3$, $X(a)|\psi\rangle=|\psi+a\rangle$, $Z(b)|\psi\rangle=(-1)^{b\cdot \psi}|\psi\rangle$, and for $a,b\in\mathbb{F}_2^n$. It is easy to verify that
$\textrm{bl}_Q(E)=\textrm{bl}_Q(\overline{E})=\textrm{bl}_s((a|b))$.
Then we have the  group framework for quantum burst error correction codes.
\begin{theorems}
\label{theorem_symplectic}
Suppose that there exists an $(n-k)$-dimensional linear subspace $C$ of $\mathbb{F}_2^{2n}$ which is contained in its symplectic dual $C^{\bot_s}$, i.e., $C\subseteq C^{\bot_s}$. Let $\ell$ be the largest integer such that for  arbitrary two vectors $e_1\neq e_2\in\mathbb{F}_2^{2n}$ whose  symplectic burst length $\leq \ell$ there is  $e_1+e_2\notin C^{\bot_s} \backslash  C$. Then there exists a quantum burst error correction code $Q=[[n,k]]$  which can correct arbitrary quantum burst errors of length $\ell$ or less. If all the $e_1+e_2\notin C^{\bot_s} \backslash  \{0\}$, then $Q$ is a nondegenerate  quantum burst error correction code.
\end{theorems}
\begin{IEEEproof}
The existence of the quantum code $Q$ has been shown in
 \cite[Theorem 1]{calderbank1998quantum}. The burst error correction abilities of  $Q$ can be obtained directly by combining
 \cite[Lemma 1]{calderbank1998quantum} and Proposition \ref{proposition_burst}.
\end{IEEEproof}

As shown in \cite{calderbank1998quantum},  binary quantum codes can be  constructed by using additive codes over $\mathbb{F}_4$. Define the
\emph{trace inner product} of two vectors $u,v\in \mathbb{F}_4^n$ by
\begin{equation}
(u,v)_{tr}= \sum_{i=1}^{n}(u_iv_j^2+u_i^2v_j).
\end{equation}
Let $C$ be an additive code over $\mathbb{F}_4$, then the \emph{trace dual} of $C$ with respect to the trace inner product is defined by
\begin{equation}
 C^{\bot_{tr}}=\{v\in \mathbb{F}_4^{n}|\forall u\in C,(u,v)_{tr}=0\}.
\end{equation}
Then Theorem \ref{theorem_symplectic} can also be reformulated by using additive codes over $\mathbb{F}_4$ and by replacing  the symplectic inner product with trace inner product.
\begin{theorems}
\label{Hermitian QBECCs}
Suppose that $C$ is  an additive code over $\mathbb{F}_4$ which is contained in its trace dual $C^{\bot_{tr}}$, i.e., $C\subseteq C^{\bot_{tr}}$.  Let $\ell$ be the largest integer such that for  arbitrary two vectors $e_1\neq e_2\in\mathbb{F}_4^{n}$ whose burst length $\leq \ell$ there is  $e_1+e_2\notin C^{\bot_{tr}} \backslash  C$. Then there exists a binary quantum burst error correction code $Q=[[n,k]]$  which can correct arbitrary quantum burst errors of length $\ell$ or less. If all the $e_1+e_2\notin C^{\bot_{tr}} \backslash  \{0\}$, then $Q$ is a nondegenerate  quantum burst error correction code.
\end{theorems}

The CSS code construction \cite{calderbank1996good,steane1996error} provides a direct way to construct QECCs from classical linear codes. The CSS construction for QBECCs can be obtained from Theorem \ref{theorem_symplectic}.

\begin{corollarys}[CSS Construction]
\label{CSS QBECCs}
Let $C_1=[n,k_1]$ and $C_2=[n,k_2]$ be
two binary linear codes which have  $\ell_1$ and $\ell_2$ burst error correction abilities, respectively,
and such that $C_2^\bot\subseteq C_1$. Let $\ell$ be the largest integer such that for arbitrary two vectors $e_1\neq e_2\in\mathbb{F}_2^{n}$ whose burst length $\leq \ell$ there is  $e_1+e_2\notin (C_1 \backslash  C_2^\bot)\cup (C_2 \backslash  C_1^\bot)$.
 Then there exists a binary quantum burst error correction code $Q=[[n,k_1+k_2-n]]$  which can correct arbitrary quantum burst errors of length $\ell$ or less and if    $\ell=\min\{\ell_1,\ell_2\}$, then $Q$ is a nondegenerate code.
\end{corollarys}

\subsection{Quantum Reiger Bound}

For a classical code  $C=[n,k]$  which can correct $\leq t$   random errors or can correct any burst errors of length $\leq \ell$,   there exists two important  upper bounds called the Singleton bound $n-k\geq 2t$ and the Reiger bound $n-k\geq 2\ell$ that constrain the random error correction and burst error correction abilities of $C$, respectively (see \cite{lin2004error}).   In quantum codes, let  $Q=[[n,k]]$ be a QECC which can correct $\leq t$ quantum random errors, there exists the quantum Singleton bound  $n-k\geq 4t$  which  is an upper bound for the quantum random error correction ability of code $Q$ (see \cite{nielsen2000quantum,calderbank1998quantum}).

In the following, we derive the quantum Reiger bound (QRB) which is   an upper bound for the quantum burst error correction ability of code $Q$.
\begin{theorems} [Quantum Reiger Bound]
\label{quantumReigerbound}
If an $[[n,k]]$ QBECC $Q$ can correct quantum burst errors of length $\ell$, then it satisfies
\begin{equation}
n-k\geq 4\ell.
\end{equation}
\end{theorems}



\begin{IEEEproof}
The proof follows closely by that of the quantum Singleton bound given by Preskill (see \cite[p.32]{preskill1998physics} and \cite[p.568]{nielsen2000quantum}).

First of all, Lemma~\ref{no-cloning bound for QBECCs} in the Appendix says that if $Q$ can correct $\ell$ burst errors, then it must satisfy $n>4\ell$, a consequence following from the quantum no-cloning principle.

Then we introduce a $k$-qubit ancilla system $A$, and construct a pure state $|\Psi\rangle_{AQ}$ that is maximally entangled between the system $A$ and the $2^k$ codewords of the $[[n,k]]$ QBECC $Q$:
\begin{equation}
|\Psi\rangle_{AQ}=\frac{1}{\sqrt{2^k}}\sum|x\rangle_A|{x}\rangle_Q,
\end{equation}
where $\{|x\rangle_A\}$ denotes an orthonormal basis for the $2^k$-dimensional Hilbert space of the ancilla, and $\{|x\rangle_Q\}$ denotes an orthonormal basis for the $2^k$-dimensional code subspace. 
It is obvious that
\begin{equation}
S(A)_\Psi=k=S(Q)_\Psi,
\end{equation}
where $S(A)_\rho= -\text{Tr} \rho_A \log \rho_A $ is the von Neumann entropy of a density operator $\rho_A$.


Next we divide the $n$-qubit QBECC $Q $ into three disjoint parts so that  $Q^{(1)}$ and $Q^{(2)}$ consist of $2\ell$ qubits each and $Q^{(3)}$ consists of the remaining $n-4\ell$ qubits. If we trace out $Q^{(2)}$ and $Q^{(3)}$, the reduced density matrix that we obtained must contain no correlations between $Q^{(1)}$ and the ancilla $A$, a consequence following from Lemma \ref{Located errors for QBECCs} in the Appendix. This means that the entropy of
system $AQ^{(1)}$ is additive:
\begin{equation}
S( Q^{(2)}Q^{(3)})_\Psi=S(AQ^{(1)})_\Psi=S(A)_\Psi+S(Q^{(1)})_\Psi.
\end{equation}
Similarly,
\begin{equation}
S( Q^{(1)}Q^{(3)})_\Psi=S(AQ^{(2)})_\Psi=S(A)_\Psi+S(Q^{(2)})_\Psi.
\end{equation}
Furthermore, in general, the von Neumann entropy is subadditive, so that
\begin{eqnarray}
S( Q^{(1)}Q^{(3)})_\Psi\leq S(Q^{(1)})_\Psi +S(Q^{(3)})_\Psi\\
S( Q^{(2)}Q^{(3)})_\Psi\leq S(Q^{(2)})_\Psi +S(Q^{(3)})_\Psi.
\end{eqnarray}
Combining these inequalities with the equalities above, we find
\begin{eqnarray}
S(A)+S(Q^{(2)})_\Psi\leq S(Q^{(1)})_\Psi +S(Q^{(3)})_\Psi\\
S(A)+S(Q^{(1)})_\Psi\leq S(Q^{(2)})_\Psi +S(Q^{(3)})_\Psi.
\end{eqnarray}
Both inequalities can be simultaneously satisfied only if
\begin{equation}
S(A)_\Psi\leq S(Q^{(3)})_\Psi.
\end{equation}
Finally, we have
\begin{equation}
S(A)_\Psi=k\leq n-4\ell,
\end{equation}
since $S(Q^{(3)})$ is bounded above by its dimension $n - 4\ell$. We then conclude the quantum Reiger bound.
\end{IEEEproof}


\section{Construction of Quantum Burst Error Correction Codes}


In this section we provide two methods for constructing QBECCs: one by using computer search based on the stabilizer formalism, and the other by concatenating and interleaving quantum tensor product codes.

\subsection{Stabilizer QBECCs Constructed by Using Computer Search}



We create a program using Magma software (version V2.12-16) to search all possible cyclic codes to a reasonable length ($n\leq 41$) according to Proposition \ref{proposition_burst}  and make the codes as
close as possible to the quantum Reiger bound. For simplicity,  we only consider the construction of QBECCs from cyclic codes with odd length.  We list new codes that are near or saturate the quantum Reiger bound found by this program in Table \ref{Computer Searching for QBECCs}.
The bold numbers ``$\textbf{1}-\textbf{3}$'' stand for the
coefficients and the superscript numbers stand for the exponents in the generator polynomials of the corresponding classical cyclic codes \cite{lin2004error}. Notice that the burst error-correcting abilities of most QBECCs in Table \ref{Computer Searching for QBECCs} can saturate the quantum Reiger bound. In particular, we get several \emph{degenerate} QBECCs which are the first class of QBECCs until now and they can  saturate the quantum Reiger bound. Moreover,  some of the constructed QBECCs are better than the QBECCs in  Ref.~\cite{tokiwa2005some}, {e.g., the codes $[[15,3]]$,   $[[35,17]]$ and  $[[35,13]]$ in Table \ref{Computer Searching for QBECCs} have larger dimensions than the codes $[[15,2]]$,   $[[35,14]]$ and  $[[35,11]]$ in Ref.~\cite{tokiwa2005some},  respectively, but have the same burst error-correcting ability.}

We remark that some QBECCs of length up to 51 have been found by using computer search in Ref.~\cite{tokiwa2005some}. However, only CSS type QBECCs were considered  and no degenerate codes were obtained in Ref.~\cite{tokiwa2005some}.

\newcommand{\tabincell}[2]{\begin{tabular}{@{}#1@{}}#2\end{tabular}}
\begin{table}
\renewcommand{\arraystretch}{0.1}
\setlength{\tabcolsep}{2pt}
\caption{Computer Searching for Quantum Burst Error Correction Codes}
\label{Computer Searching for QBECCs}
\centering
\footnotesize
\begin{tabular}[c]{|c|c|c|c|c|}
\hline

 $[[n,k]]$&  $l$ &Generator Polynomials&QRB&Degenerate?  \\
 \hline
 $[[13,1]]$&3&$g=(\textbf{1}^6\textbf{2}^5\textbf{3}^3\textbf{2}^1\textbf{1}^0)$&3&False\\
\hline
 $[[15,3]]$&3&$g=(\textbf{1}^6\textbf{2}^3\textbf{1}^0)$&3&False\\
\hline
$[[17,1]]$&4&$g=(\textbf{1}^8\textbf{3}^7\textbf{1}^6\textbf{1}^5\textbf{2}^4\textbf{1}^3\textbf{1}^2\textbf{3}^1\textbf{1}^0)$&4&False\\
\hline
$[[17,1]]$&4&$g=(\textbf{1}^8\textbf{3}^7\textbf{3}^5\textbf{3}^4\textbf{3}^3\textbf{3}^1\textbf{1}^
0)$&4&True\\
\hline
$[[21,9]]$&2&\tabincell{l}{$g_1=(\textbf{1}^{6}\textbf{1}^4\textbf{1}^1\textbf{1}^0)$,
\\$g_2=(\textbf{1}^{6}\textbf{1}^4\textbf{1}^2\textbf{1}^1\textbf{1}^0)$}&3&False\\
\hline
$[[23,1]]$&5&$(g=\textbf{1}^{11}\textbf{1}^9\textbf{1}^7\textbf{1}^6\textbf{1}^5\textbf{1}^1\textbf{1}^0)$&5&False\\
\hline
$[[25,1]]$&6&$g=(\textbf{1}^{12}\textbf{2}^{11}\textbf{1}^{10}\textbf{2}^7\textbf{3}^6\textbf{2}^5\textbf{1}^2\textbf{2}^1\textbf{1}^0)$&6&True\\
\hline
$[[25,5]]$&5&$g=(\textbf{1}^{10}\textbf{2}^{5}\textbf{1}^0)$&5&False\\
\hline
$[[29,1]]$&7&\tabincell{c}{$g=(\textbf{1}^{14}\textbf{2}^{13}\textbf{2}^{11}\textbf{3}^{10}\textbf{1}^{9}\textbf{3}^{8} $
\\\hspace{10mm}$\textbf{2}^7\textbf{3}^6\textbf{1}^5\textbf{3}^{4}\textbf{2}^3\textbf{2}^1\textbf{1}^0)$}&7&True\\
\hline
$[[35,25]]$&2&$g=(\textbf{1}^{5}\textbf{2}^{4}\textbf{3}^{2}\textbf{2}^{1}\textbf{1}^0)$&2&False\\
\hline
$[[35,19]]$&3&$g=(\textbf{1}^{8}\textbf{2}^{7}\textbf{3}^{6}\textbf{1}^{5}\textbf{3}^{3}\textbf{1}^{2}\textbf{1}^1\textbf{1}^0)$&4&False\\
\hline
$[[35,17]]$&4&$g=(\textbf{1}^{9}\textbf{3}^{7}\textbf{3}^{6}\textbf{3}^{5}\textbf{3}^{4}\textbf{2}^{3}\textbf{2}^{2}\textbf{2}^1\textbf{1}^0)$&4&False\\
\hline
$[[35,13]]$&5&$g=(\textbf{1}^{11}\textbf{3}^{10}\textbf{2}^{9}\textbf{1}^{8}\textbf{2}^{7}\textbf{2}^{6}\textbf{3}^{5}
\textbf{1}^{3}\textbf{2}^{2}\textbf{1}^1\textbf{1}^0)$&5&False\\
\hline
$[[35,7]]$&6&\tabincell{c}{$g=(\textbf{1}^{14}\textbf{2}^{13}\textbf{3}^{11}\textbf{2}^{10}\textbf{2}^{9}\textbf{3}^{8} $
\\\hspace{10mm}$\textbf{1}^7\textbf{1}^6\textbf{2}^5\textbf{3}^{3}\textbf{2}^2\textbf{3}^1\textbf{1}^0)$}&7&False\\
\hline
$[[41,1]]$&10&\tabincell{c}{$g_1=(\textbf{1}^{20}\textbf{1}^{18}\textbf{1}^{17}\textbf{1}^{16}\textbf{1}^{15}\textbf{1}^{14}\textbf{1}^{11}$
\\\hspace{10mm}$\textbf{1}^{10}\textbf{1}^{9}\textbf{1}^{6}\textbf{1}^{5}\textbf{1}^{4}\textbf{1}^{3}\textbf{1}^{2}\textbf{1}^{0})$,\\
$g_2=(\textbf{1}^{20}\textbf{1}^{19}\textbf{1}^{17}\textbf{1}^{16}\textbf{1}^{14}\textbf{1}^{11}$\\
\hspace{10mm}$\textbf{1}^{10}\textbf{1}^{9}\textbf{1}^{6}\textbf{1}^{4}\textbf{1}^{3}\textbf{1}^{1}\textbf{1}^{0})$}&10&True\\
\hline
\end{tabular}
\end{table}

\subsection{Concatenation Construction of QBECCs Based on Quantum Tensor Product Codes}
In this section we give a concatenation construction of long  QBECCs from two short component codes based on the  quantum tensor product codes structure \cite{jihao2017on,fan2017comments}.

Firstly we present a brief review of classical and quantum tensor product codes. Details could be found in, e.g.,   \cite{wolf1965codes,jihao2017on}.

Let  $C_1=[n_1, k_1]_2$ be a   linear code with a
parity check matrix $H_1$ and let $\rho_1=n_1-k_1$ be the number of check symbols. Let $C_2=[n_2, k_1]_{2^{\rho_1}} $ be a  linear code over the extension filed $\mathbb{F}_{2^{\rho_1}}$ with 
a parity check matrix $H_2$. Then  the tensor product code (TPC) of $C_1$ and $C_2$ is denoted by $\mathcal{C}=C_2\otimes_HC_1$, and the parity check
matrix $\mathcal{C}$ is given by
\begin{equation}
H=H_2\otimes H_1.
\end{equation}
By selecting different types of  component codes, TPCs   can be designed to provide different error control abilities.

In \cite{jihao2017on}, a framework for the construction of quantum tensor product codes (QTPC), which can provide a wide variety of  quantum error-correcting, error-detecting or error-locating
properties, was proposed. In particular,  if one of the
component codes is selected as a BECC, then QTPCs can have  multiple quantum burst error-correcting abilities, but provided these bursts fall in distinct subblocks.
\begin{theorems}[\cite{jihao2017on}]
\label{CSS_TPC1_theorem}
Let $C_1=[n_1,k_1]_q$ be an  $\ell_1$ burst error correction code, and let $C_2=[n_2,k_2]_{q^{\rho_1}}$ be  an  $\ell_2$ burst error correction code
over the extension field $\mathbb{F}_{q^{\rho_1}}$, and the numbers of   check symbols are $\rho_1=n_1-k_1$ and $\rho_2=n_2-k_2$, respectively.
If $q=2$ and $C_1^\perp\subseteq C_1$  or
if $q=4$ and  $C_1^{\perp_h}\subseteq C_1$, where   $C_1^{\perp_h}$ is the Hermitian dual code of $C_1$, then there exists a   QTPC $\mathcal{Q}=[[n_1n_2, n_1n_2-2\rho_1\rho_2]]$ which can correct $\ell_2$ or fewer bursts of burst errors each  is a burst of length   $\ell_1$ or less,
 provided these bursts fall in distinct subblocks.
\end{theorems}

Although we can use QTPCs to correct a single burst of errors since QTPCs have multiple burst error correction abilities, they are not efficient enough any more. To overcome this problem, we can  interleave the encoded qubits before sending into the quantum channels, and deinterleave after receiving the qubits.
The whole interleaving/deinterleaving procedure is summarized as follows:
\begin{itemize}
\item[(1).] After the quantum encoding, we arrange the encoded $n_1n_2$  qubits into an $n_1\times n_2$ code array.
\item[(2).] Instead of transmitting the encoded qubits sequentially one by one, we do an interleaved transmission. Denote by $\ell_1$ the burst error correction ability of $C_1$.
If $\ell_1|n_1$ and
the component code $C_2$ can correct \emph{end-around} (see \cite{lin2004error}) burst errors, then we divide the  $n_1\times n_2$ code array into $s=n_1/\ell_1$ subblocks by rows.
We do the transmission subblock by subblock, and in each subblock, we transmit the qubits column by column sequentially (each column contains $\ell_1$ qubits).
\item[(3).] After receiving all the $n_1n_2$ qubits, we deinterleave the qubits into an $n_1\times n_2$ code array so that the quantum decoding can be processed next,
and the deinterleaving is just the inverse of the interleaving. The deinterleaving/interleaving procedure can be accomplished by using quantum SWAP gates (see \cite{nielsen2000quantum}).
\end{itemize}
\begin{figure}[!t]
\centering
\includegraphics[width=3.2in]{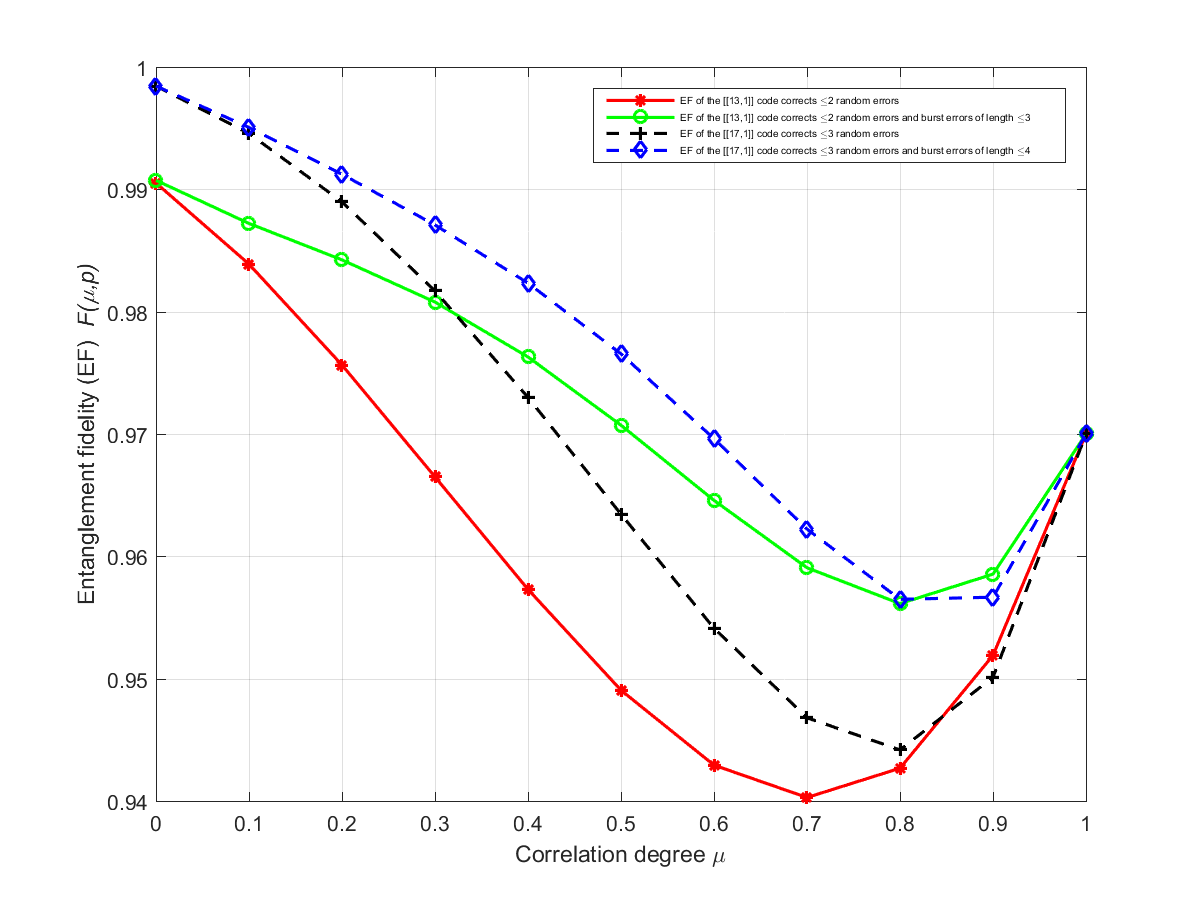}
\caption{The entanglement fidelity (EF) of the two $[[13,1]]$ and $[[17,1]]$ codes with respect to the correlation degree $  0\leq \mu\leq1 $, the error probability
is set to be $p=3\times 10^{-2}$.}
\label{ef1}
\end{figure}
Suppose that a single burst errors of length at most $\ell_1\ell_2$ happens among the $n_1n_2$ interleaved qubits. After the quantum transmition and   deinterleaving, the $n_1n_2$ qubits are recovered to their original positions,  but the single burst errors of length at most  $\ell_1\ell_2$ has been dispersed into    $\ell_2$ or fewer consecutive subblocks (end around) and each subblock  contains  a burst errors of length at most $\ell_1$.  Thus the resultant QTPC $\mathcal{Q}$ can correct a single burst error of length at most
$\ell_1\ell_2$  according to Theorem \ref{CSS_TPC1_theorem} and Ref. ~\cite{jihao2017on}. Then we have the following result.
\begin{theorems}
\label{interleav burst error correction}
Let $C_1=[n_1,k_1]_q$ and $C_2=[n_2,k_2]_{q^{\rho_1}}$ be two component codes of a QTPC with parameters
$\mathcal{Q}=[[n_1n_2,n_1n_2-2\rho_1\rho_2]]$, where $C_1$ is an $\ell_1$ burst error correction code and
$C_2$ is an $\ell_2$ burst error correction code (end-around),
and the numbers of check symbols are $\rho_1=n_1-k_1$ and $\rho_2=n_2-k_2$, respectively. If $\ell_1|n_1$,
then there exists an   $\ell_1\ell_2$ burst error correction quantum code $\mathcal{Q}=[[n_1n_2,n_1n_2-2\rho_1\rho_2]]$.
\end{theorems}
\begin{examples}
\label{interleav burst error correction corollarys}
We choose  $C_1=[15,9]_4$ as a $3$ burst error correction cyclic code with the generator polynomial $g=(\textbf{1}^6\textbf{2}^3\textbf{1}^0)$ and it is Hermitian dual containing by Table \ref{Computer Searching for QBECCs}. Let $C_2=[n_2,n_2-2\ell_2,2\ell_2+1]_{4^{6}}$ be an MDS code over the extension field $\mathbb{F}_{4^{6}}$ with $2\leq n_2\leq 4^6+2$ and $1\leq \ell_2\leq\lfloor\frac{n_2-1}{2}\rfloor$. Then there exists a $3\ell_2$ burst error correction QTPC with parameters $\mathcal{Q}=[[15n_2,15n_2-24\ell_2]]$.
\end{examples}

\begin{figure}[!t]
\centering
\includegraphics[width=3.2in]{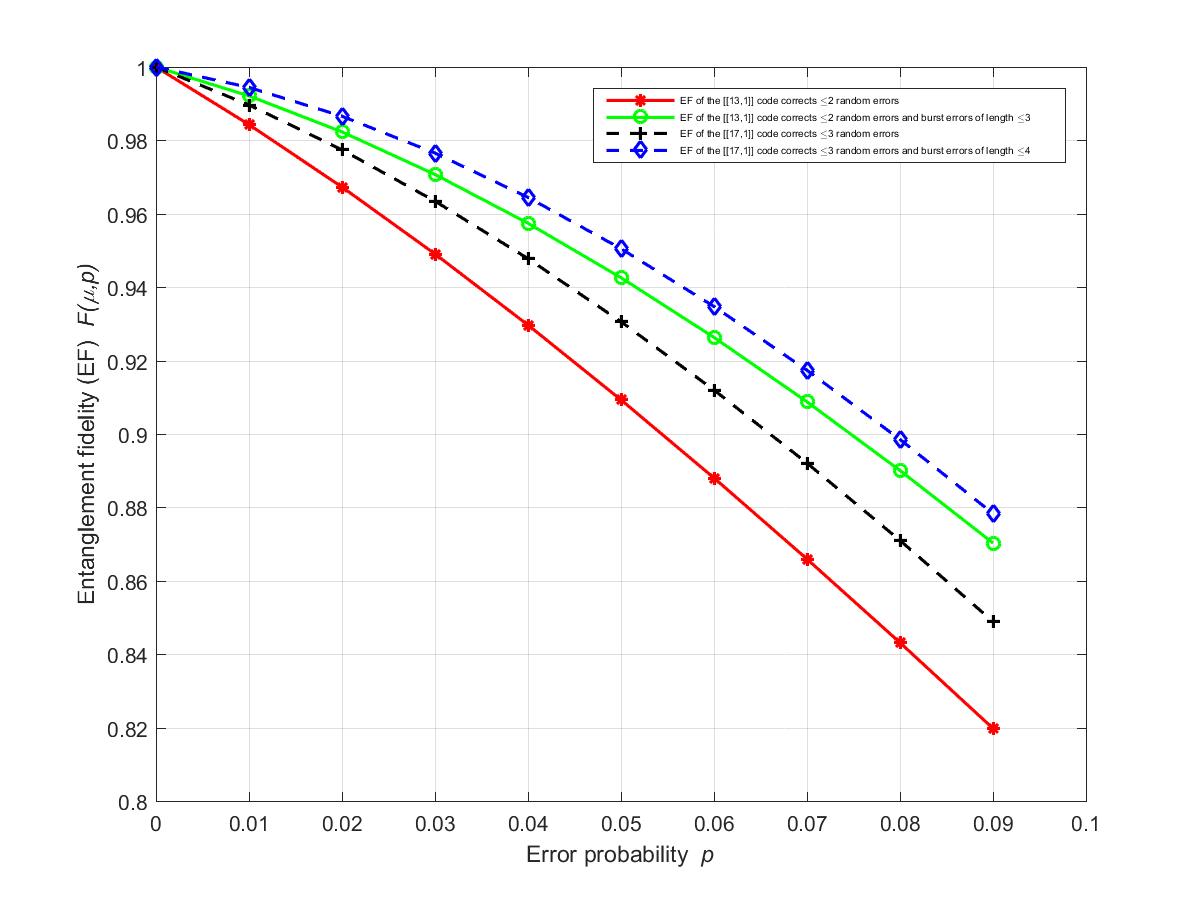}
\caption{ The entanglement fidelity (EF) of the two $[[13,1]]$ and $[[17,1]]$ codes with respect to the error probability
 $ 1\times 10^{-5}\leq p\leq1\times10^{-1}$, the correlation degree is
  set to be $\mu=0.5$.}
\label{ef2}
\end{figure}

\section{Performance of QBECCs over Markovian Correlated Quantum Memory Channels}
In this section we evaluate  the performance of two specific QBECCs in the presence of correlated errors.

The channel model that we choose is a Markovian correlated   depolarizing quantum channel \cite{caruso2014quantum,Cafaro2010quantum}: 
\begin{eqnarray}
  \Phi^{(n)}(\rho) =\sum_{i_1,\ldots,i_n=0}^{3}p_{i_n|i_{n-1}}^{(n)}\cdots p_{i_2|i_{1}}^{(2)}p_{i_1}^{(1)}\nonumber\\
\times  E_{i_n}^{(n)}\cdots E_{i_1}^{(1)}\rho E_{i_1}^{(1)\dagger}\cdots E_{i_n}^{(n)\dagger}, 
\end{eqnarray}
where $\{E_{i_j}^{(j)}\}_{i,j}$ are the Pauli operators, and the conditional probabilities
 satisfy the normalization condition
\begin{equation}
\sum\limits_{i_1,\ldots,i_n=0}^{3}p_{i_n|i_{n-1}}^{(n)}\cdots p_{i_2|i_{1}}^{(2)}p_{i_1}^{(1)}=1,
\end{equation}
where $p_{l|k}^{(j)}=(1-\mu)p_{l}+\mu\delta_{(k,l)}$ for $1\leq j\leq n$ and $ 0\leq k,l\leq 3$, and $p_{0}=1-p, p_{1,2,3}=p/3$
are the error probabilities in the  depolarizing channel, $\mu\in[0,1]$ is the correlation degree.

Specifically, we show the performance of the two specific codes by measuring the  entanglement fidelity  $\mathcal{F}(p,\mu)$ as
a  function of the error probability $p$ and the  correlation degree $\mu$ \cite{nielsen2000quantum}. 

The two specific codes considered here are two QBECCs  $[[13,1]]$   and    $[[17,1]]$   in Table \ref{Computer Searching for QBECCs} which can correct burst errors of length $\leq 3$ and of length $\leq 4$, respectively. Through the computation, we know that the minimum distances of the two codes are $5$ and $7$, then they can also  correct $\leq2$ and $\leq3$ random errors, respectively, and they have achieved the upper bounds in Ref.~\cite{Grassl:codetables}. We plot the performance by means of entanglement fidelity of the two specific codes with respect to  random errors or burst errors,  versus the correlation degree $\mu$ or the error probability $p$  in Fig. \ref{ef1} or Fig. \ref{ef2}, respectively. For details about
the computation of the entanglement fidelity, see \cite{caruso2014quantum,Cafaro2010quantum,nielsen2000quantum}, and the computation results are put in the \href{https://www.dropbox.com/sh/on17nv0ca1b6j7j/AAD3fd0PORxyQH0TuhJlJR0ra?dl=0}{Cloud}.
\hspace{-2mm}

\begin{figure}[!t]
\centering
\includegraphics[width=3.2in]{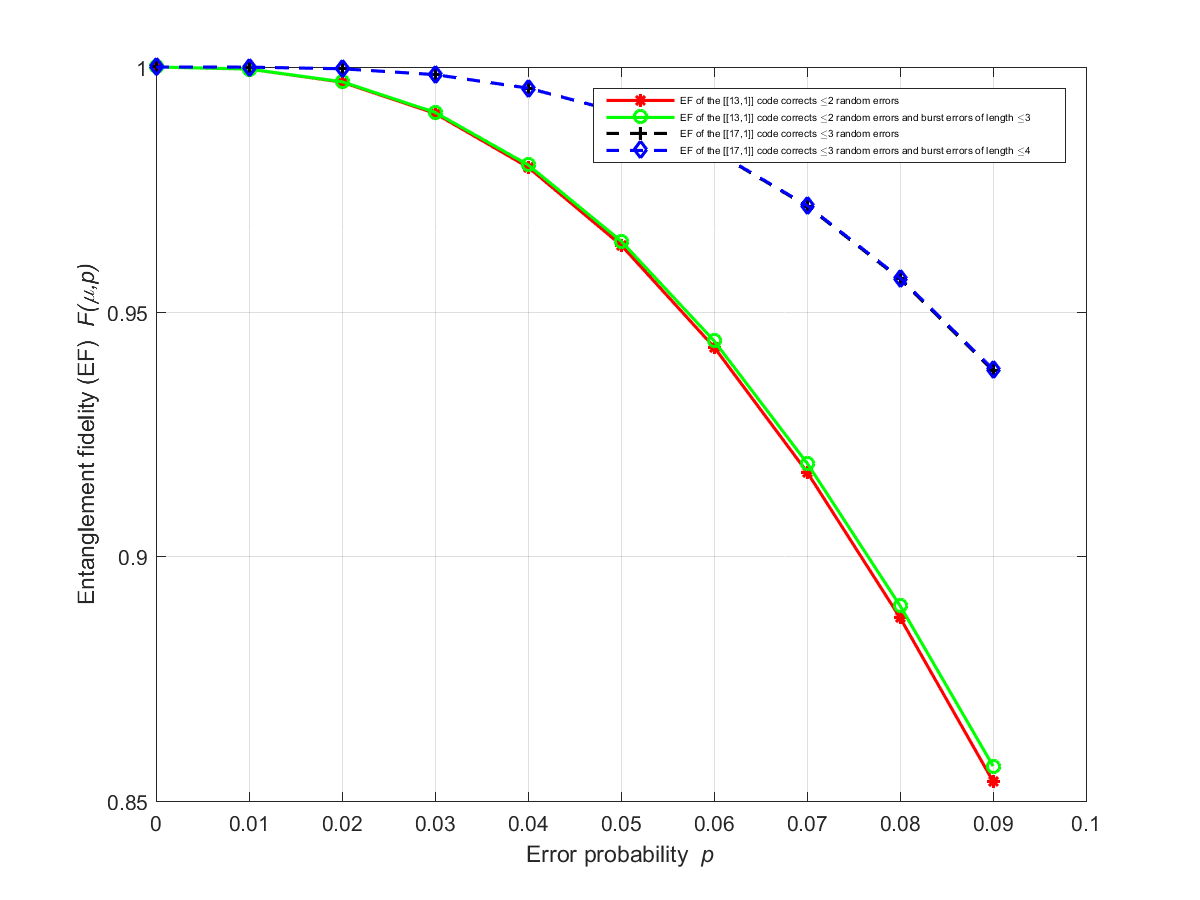}
\caption{ The entanglement fidelity (EF) of the two $[[13,1]]$ and $[[17,1]]$ codes with respect to the error probability
 $ 1\times 10^{-5}\leq p\leq1\times10^{-1}$, the correlation degree is
  set to be $\mu=0$.}
\label{ef3}
\end{figure}

It is shown that the correlation errors do degrade the performance of the entanglement fidelity of the two codes in Fig.~\ref{ef1}. If the the correlation degree $\mu=0$ which means that errors are independent with each other, then the extra burst error correction abilities of the two codes do little help to improve the performance of entanglement fidelity of them, respectively, see Fig. \ref{ef1} and Fig. \ref{ef3}.  However, if we consider the correlated errors when $0<\mu<1$, the performance of the entanglement fidelity can be improved largely, see Fig.~\ref{ef1} and Fig.~\ref{ef2}. In particular, in Fig.~\ref{ef2}, the $[[13,1]]$ code have  better performance when considering its extra burst error correction ability compared to the  $[[17,1]]$ code when only considering the
random error correction ability.

\section*{Acknowledgment}

J. Fan was supported by  NJIT (Grant No. YKJ201719) and by NSFC (Grant No.  61403188). M.-H. Hsieh was supported in part by an ARC Future Fellowship under Grant FT140100574 and in part by U.S. the Army Research Office for Basic Scientific Research under Grant W911NF-17-1-0401.

\ifCLASSOPTIONcaptionsoff
  \newpage
\fi



%
\bibliographystyle{IEEEtran}
\bibliography{IEEEabrv,bibfileISIT2018QBECC}

\begin{thebibliography}{10}
\providecommand{\url}[1]{#1}
\csname url@samestyle\endcsname
\providecommand{\newblock}{\relax}
\providecommand{\bibinfo}[2]{#2}
\providecommand{\BIBentrySTDinterwordspacing}{\spaceskip=0pt\relax}
\providecommand{\BIBentryALTinterwordstretchfactor}{4}
\providecommand{\BIBentryALTinterwordspacing}{\spaceskip=\fontdimen2\font plus
\BIBentryALTinterwordstretchfactor\fontdimen3\font minus
  \fontdimen4\font\relax}
\providecommand{\BIBforeignlanguage}[2]{{%
\expandafter\ifx\csname l@#1\endcsname\relax
\typeout{** WARNING: IEEEtran.bst: No hyphenation pattern has been}%
\typeout{** loaded for the language `#1'. Using the pattern for}%
\typeout{** the default language instead.}%
\else
\language=\csname l@#1\endcsname
\fi
#2}}
\providecommand{\BIBdecl}{\relax}
\BIBdecl

\bibitem{lin2004error}
S.~Lin and D.~J. Costello, \emph{Error Control Coding: Fundamentals and
  Applications}, 2nd~ed.\hskip 1em plus 0.5em minus 0.4em\relax Upper Saddle
  River, NJ: Prentice-Hall, Inc., 2004.

\bibitem{steane1996error}
A.~M. Steane, ``Error correcting codes in quantum theory,'' \emph{Phys. Rev.
  Lett.}, vol.~77, no.~5, p. 793, 1996.

\bibitem{calderbank1996good}
A.~R. Calderbank and P.~W. Shor, ``Good quantum error-correcting codes exist,''
  \emph{Phys. Rev. A}, vol.~54, no.~2, p. 1098, 1996.

\bibitem{calderbank1997quantum}
A.~R. Calderbank, E.~M. Rains, P.~W. Shor, and N.~J. Sloane, ``Quantum error
  correction and orthogonal geometry,'' \emph{Phys. Rev. Lett.}, vol.~78,
  no.~3, p. 405, 1997.

\bibitem{Brun436}
T.~Brun, I.~Devetak, and M.-H. Hsieh, ``Correcting quantum errors with
  entanglement,'' \emph{Science}, vol. 314, no. 5798, pp. 436--439, 2006.

\bibitem{6671483}
M.~M. Wilde, M.~H. Hsieh, and Z.~Babar, ``Entanglement-assisted quantum turbo
  codes,'' \emph{IEEE Trans. Inf. Theory}, vol.~60, no.~2, pp. 1203--1222, Feb
  2014.

\bibitem{jihao2017on}
J.~Fan, Y.~Li, M.-H. Hsieh, and H.~Chen, ``On quantum tensor product codes,''
  \emph{Quantum Inf. Comput.}, vol.~17, no. 13$\&$14, pp. 1105--1122, 2017.

\bibitem{vatan1999spatially}
F.~Vatan, V.~P. Roychowdhury, and M.~Anantram, ``Spatially correlated qubit
  errors and burst-correcting quantum codes,'' \emph{IEEE Trans. Inf. Theory},
  vol.~45, no.~5, pp. 1703--1708, 1999.

\bibitem{kawabata2000quantum}
S.~Kawabata, ``Quantum interleaver: quantum error correction for burst error,''
  \emph{J. Phys. Soc. Jpn.}, vol.~69, no.~11, pp. 3540--3543, 2000.

\bibitem{tokiwa2005some}
K.-i. Tokiwa, K.~Kiyama, T.~Yamasaki \emph{et~al.}, ``Some binary quantum codes
  with good burst-error-correcting capabilities,'' 2005.

\bibitem{PhysRevA.72.062323}
D.~Kretschmann and R.~F. Werner, ``Quantum channels with memory,'' \emph{Phys.
  Rev. A}, vol.~72, p. 062323, Dec 2005.

\bibitem{caruso2014quantum}
F.~Caruso, V.~Giovannetti, C.~Lupo, and S.~Mancini, ``Quantum channels and
  memory effects,'' \emph{Rev. Mod. Phys.}, vol.~86, no.~4, p. 1203, 2014.

\bibitem{7501532}
C.~Y. Lai, M.~H. Hsieh, and H.~F. Lu, ``On the macwilliams identity for
  classical and quantum convolutional codes,'' \emph{IEEE Trans. Commun.},
  vol.~64, no.~8, pp. 3148--3159, Aug 2016.

\bibitem{6461941}
Y.~Fujiwara and V.~D. Tonchev, ``A characterization of entanglement-assisted
  quantum low-density parity-check codes,'' \emph{IEEE Trans. Inf. Theory},
  vol.~59, no.~6, pp. 3347--3353, June 2013.

\bibitem{5714249}
M.~H. Hsieh, W.~T. Yen, and L.~Y. Hsu, ``High performance entanglement-assisted
  quantum ldpc codes need little entanglement,'' \emph{IEEE Trans. Inf.
  Theory}, vol.~57, no.~3, pp. 1761--1769, March 2011.

\bibitem{PhysRevA.79.032340}
M.-H. Hsieh, T.~A. Brun, and I.~Devetak, ``Entanglement-assisted quantum
  quasicyclic low-density parity-check codes,'' \emph{Phys. Rev. A}, vol.~79,
  p. 032340, Mar 2009.

\bibitem{PhysRevA.76.062313}
M.-H. Hsieh, I.~Devetak, and T.~Brun, ``General entanglement-assisted quantum
  error-correcting codes,'' \emph{Phys. Rev. A}, vol.~76, p. 062313, Dec 2007.

\bibitem{Cafaro2010quantum}
C.~Cafaro and S.~Mancini, ``Quantum stabilizer codes for correlated and
  asymmetric depolarizing errors,'' \emph{Phys. Rev. A}, vol.~82, no.~1, p.
  012306, 2010.

\bibitem{nielsen2000quantum}
M.~A. Nielsen and I.~L. Chuang, \emph{Quantum Computation and Quantum
  Information}.\hskip 1em plus 0.5em minus 0.4em\relax Cambridge University
  Press, 2000, no.~2.

\bibitem{6778074}
T.~A. Brun, I.~Devetak, and M.~H. Hsieh, ``Catalytic quantum error
  correction,'' \emph{IEEE Trans. Inf. Theory}, vol.~60, no.~6, pp. 3073--3089,
  June 2014.

\bibitem{bennett1996mixed}
C.~H. Bennett, D.~P. DiVincenzo, J.~A. Smolin, and W.~K. Wootters,
  ``Mixed-state entanglement and quantum error correction,'' \emph{Phys. Rev.
  A}, vol.~54, no.~5, p. 3824, 1996.

\bibitem{knill1997theory}
E.~Knill and R.~Laflamme, ``Theory of quantum error-correcting codes,''
  \emph{Phys. Rev. A}, vol.~55, no.~2, p. 900, 1997.

\bibitem{calderbank1998quantum}
A.~Calderbank, E.~Rains, P.~Shor, and N.~Sloane, ``Quantum error correction via
  codes over {GF(4)},'' \emph{IEEE Trans. Inf. Theory}, vol.~44, no.~4, pp.
  1369--1387, 1998.

\bibitem{preskill1998physics}
J.~Preskill, ``Physics 229: Advanced mathematical methods of physics--quantum
  computation and information,'' \emph{California Institute of Technology},
  1998.

\bibitem{fan2017comments}
J.~Fan and H.~Chen, ``Comments on and corrections to ``on the equivalence of
  generalized concatenated codes and generalized error location codes'',''
  \emph{IEEE Trans. Inf. Theory}, vol.~63, no.~8, pp. 5437--5439, 2017.

\bibitem{wolf1965codes}
J.~K. Wolf, ``On codes derivable from the tensor product of check matrices,''
  \emph{IEEE Trans. Inf. Theory}, vol.~11, no.~2, pp. 281--284, 1965.

\bibitem{Grassl:codetables}
M.~Grassl, ``Bounds on the minimum distance of linear codes and quantum
  codes,'' Online available at \url{http://www.codetables.de}, 2007.

\end{thebibliography}
%








\appendix

 \begin{lemmas}[No-Cloning Bound]
  \label{no-cloning bound for QBECCs}
  For an arbitrary $\ell$ burst error correction code $C=[[n,k\geq 1]]$   exists only
  if
  \begin{equation}
  n>4\ell.
  \end{equation}
  \end{lemmas}
\begin{IEEEproof}
Suppose that there exists a code  $C'=[[n,k\geq 1]]$ with $2\leq n\leq4\ell$. After encoding $k$ qubits into $n$ ones, we split the
encoded block into two sub-blocks, one contains the first $\lfloor \frac{n}{2}\rfloor$ qubits and the other contains the rest of the
$n-\lfloor \frac{n}{2}\rfloor$ qubits.

If we append $\lfloor \frac{n}{2}\rfloor$  ancilla qubits $|0\cdots0\rangle$  to the first sub-block, and
append $n-\lfloor \frac{n}{2}\rfloor$  ancilla qubits $|0\cdots0\rangle$  to the second sub-block, then the original encoded
block has spawned two offspring, the first one with  located burst errors of length
at most $\lfloor \frac{n}{2}\rfloor$, and the second one with located burst errors of length
at most  $n-\lfloor \frac{n}{2}\rfloor$. If we were able to
correct the two located burst errors in each of the offspring (see Lemma \ref{Located errors for QBECCs}), we would obtain two
identical copies of the parent encoded block, which is a contradiction with the quantum no-cloning theorem \cite{nielsen2000quantum}. Therefor we must have $n>4\ell.$
\end{IEEEproof}

\begin{lemmas}[Located Burst Errors]
\label{Located errors for QBECCs}
For a  QECC $Q=[[n,k]]$ that corrects arbitrary burst errors of length  $\ell$ or less can correct  located burst errors of length at most $2\ell$.
\end{lemmas}

    \begin{IEEEproof}
  Denote   an arbitrary error of length $n$ by
  \begin{equation}
  e= e_1\otimes\ldots\otimes e_x\otimes\ldots\otimes e_y\otimes\ldots\otimes e_n ,
  \end{equation}
where $1\leq x<y\leq n, y-x+1=2\ell$, and $e_i(1\leq i\leq n)$ are Pauli matrices.
The set $E(x,y)$ of burst errors to be corrected is the set of all Pauli operators, {where each acts trivially on the qubits $1$ to $x-1$ and on the qubits $y+1$ to $n$ (except $x=1$ and $y=n$)}. Then each error  in $E(x,y)$ has a burst length of at most $2\ell$.
But now, for each $E_a$ and $E_b$ in $E(x,y)$, the product $E_a^\dagger E_b$
also has a burst of length at most $2\ell$. Therefore, the burst error-correcting criterion (\ref{burst-error-correction criterion})
is satisfied for all $E_{a,b}\in E$, provided $Q$ is an $\ell$ burst error correction code.
  \end{IEEEproof}


\end{document}